\documentclass[conference]{IEEEtran}
\IEEEoverridecommandlockouts
\usepackage{amsmath, amssymb, graphicx, cite, xcolor, algorithmic, textcomp}
\usepackage{subfig}
\usepackage{colortbl}
\usepackage{xcolor}
\usepackage{afterpage}
\usepackage{placeins}
\usepackage{subfig}
\usepackage{colortbl}
\usepackage{pgfplotstable}
\usepackage{xcolor}
\usepackage{multirow}
\usepackage{booktabs}
\usepackage{url}
\usepackage{hyperref}

\definecolor{red}{rgb}{1,0,0}
\definecolor{green}{rgb}{0,1,0}
\definecolor{blue}{rgb}{0,0,1}

\title{MS-SCANet: A Multiscale Transformer-Based Architecture with Dual Attention for No-Reference Image Quality Assessment}

\author{\IEEEauthorblockN{Mayesha Maliha R. Mithila and Myl\`ene C.Q. Farias }
\IEEEauthorblockA{\textit{Department of Computer Science} \\
\textit{Texas State University}\\
San Marcos, USA \\
elx12@txstate.edu, mylene@txstate.edu}
}

\begin{document}
%
\maketitle
%


\begin{abstract}
We present the Multi-Scale Spatial Channel Attention Network (MS-SCANet), a transformer-based architecture designed for no-reference image quality assessment (IQA). MS-SCANet features a dual-branch structure that processes images at multiple scales, effectively capturing both fine and coarse details, an improvement over traditional single-scale methods. By integrating tailored spatial and channel attention mechanisms, our model emphasizes essential features while minimizing computational complexity. A key component of MS-SCANet is its cross-branch attention mechanism, which enhances the integration of features across different scales, addressing limitations in previous approaches. We also introduce two new consistency loss functions, Cross-Branch Consistency Loss and Adaptive Pooling Consistency Loss, which maintain spatial integrity during feature scaling, outperforming conventional linear and bilinear techniques. Extensive evaluations on datasets like KonIQ-10k, LIVE, LIVE Challenge, and CSIQ show that MS-SCANet consistently surpasses state-of-the-art methods, offering a robust framework with stronger correlations with subjective human scores.
\end{abstract}


\section{Introduction}
Image quality assessment (IQA) plays a vital role in the analysis and improvement of the perceptual quality of images in numerous applications, including image compression, restoration, and enhancement. Traditional techniques, such as SSIM and PSNR, struggle to predict and generalize across various distortions~\cite{wang2004image, mittal2012making}. In the past decade, Convolutional Neural Networks (CNNs) have been frequently used in IQA applications because of their remarkable capacity to extract features crucial to visual perception \cite{kang2014convolutional}. Despite their distinctive advantages, CNNs are restricted in recognizing global contexts as they primarily concentrate on local features~\cite{zhang2018unreasonable}.

Transformers have recently proven to be particularly effective in IQA due to their ability to capture long-range dependencies via self-attention techniques.~\cite{vaswani2017attention}.  Vision Transformer (ViT) \cite{dosovitskiy2020image} demonstrates this approach by modeling the relationships between image patches as sequential data and using self-attention to achieve notable results. However, the self-attention mechanism of ViT incurs a substantial computational cost, which increases quadratically with the input size. To address these problems, hybrid models such as the Swin Transformer~\cite{liu2021swin} have been introduced, to optimize computational efficiency while capturing global context. Despite these benefits, these models struggle with complicated scenarios because their hierarchical structures may not capture all features on different scales.

CrossViT \cite{chen2021crossvit} advances feature representation by employing separate branches to process image patches of varying sizes, which are then fused using cross-attention mechanisms. Multiscale approaches, such as MAMIQA \cite{yu2023mamiqa} and the dual-branch ViT \cite{lee2023dual}, demonstrate the benefits of integrating features at multiple scales for IQA. However, some techniques such as TOPIQ \cite{chen2024topiq}, tend to treat multiscale features independently or in a bottom-up manner, which can limit the effective integration of hierarchical features.
SCA-Net \cite{roy2018scanet}, on the other hand, uses a combination of channel and spatial attention mechanisms for medical image segmentation. Similarly, the Dual Vision Transformer (DaViT) \cite{ding2021davit} uses these techniques to improve image classification.  However, the increased complexity of these models results in higher computational demands, especially for images with a high resolution.

\begin{figure*}[htbp]
\centering
\includegraphics[width=1.0\linewidth]{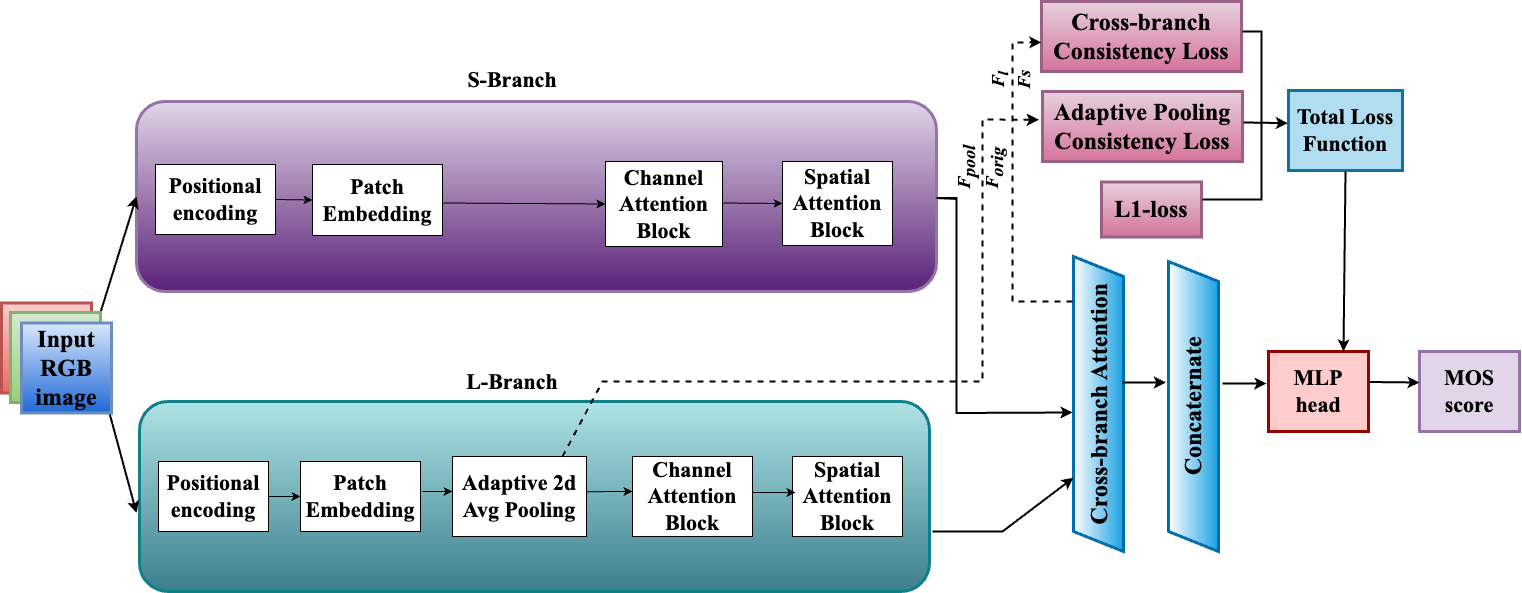}
\caption{Architecture of the proposed Multi-Scale SCANet (MS-SCANet) framework}
\label{fig:1}
\end{figure*}

To address the shortcomings of previous approaches, we present the Multi-Scale Spatial Channel Attention Network (MS-SCANet) for Blind Image Quality Assessment. Our model incorporates spatial and channel attention mechanisms into a multi-scale transformer, improving its ability to capture and emphasize key aspects at various sizes, including both local and global structures. In contrast to existing cross-attention methodologies employed for classification tokens and patch tokens, our innovative cross-branch attention mechanism is designed to integrate characteristics from different branches across multiple scales. We also propose two novel consistency loss functions for IQA: Cross-Branch Consistency Loss, which aims at ensuring consistent feature integration between different scales, and Adaptive Pooling Consistency Loss, which preserves feature integrity during downsampling. By integrating these advances, our approach demonstrates enhanced reliability in image quality assessment, supported by the comprehensive evaluations performed on IQA benchmarks. The implementation of MS-SCANet is freely accessible at \url{https://github.com/mithila442/MS-SCANet}.

\section{Proposed Model}

Figure ~\ref{fig:1} illustrates the structure of MS-SCANet, which predicts Mean Opinion Score (MOS) in the final layer. MS-SCANet employs a dual-branch architecture to handle data at many scales, which is critical to determining perceived image quality. Each branch divides the input into patches at various scales and projects them into a three-dimensional space. Positional encoding is utilized to encode spatial data.

The inputs are divided into non-overlapping windows, with self-attention applied to each segment \cite{ding2021davit}. This method reduces computational complexity by focusing on localized regions rather than deploying attention globally. Within each window $W$, the attention is calculated as follows:
\begin{equation}
\text{Att}_W(\mathbf{Q}_W, \mathbf{K}_W, \mathbf{V}_W) = \text{softmax}\left(\frac{\mathbf{Q}_W \mathbf{K}_W^T}{\sqrt{d_k}}\right) \mathbf{V}_W,
\end{equation}
The query, key, and value matrices for the window $W$ are $\mathbf{Q}_W $, $\mathbf{K}_W $, and $\mathbf{V}_W $, respectively, and the dimensionality of the key vectors is $d_k $. The attention weights are calculated using the softmax operation on the scaled dot product of $\mathbf{Q}_W $ and $\mathbf{K}_W^T $.

Using a squeeze-and-excitation network \cite{hu2018squeeze}, channel attention highlights key feature channels. This procedure re-weights the features across channels by generating an excitation map using global average pooling and convolutional layers, subsequently utilized to modulate the input feature maps. First, the squeezed feature map $\mathbf{F}_{\text{sq}} $ is computed as follows:
\begin{equation}
\mathbf{F}_{\text{sq}} = \text{Conv2D}_{1}(\text{GAP}(\mathbf{F})),
\end{equation}
where $\mathbf{F} $ is the input feature map, GAP is the global average pooling operation, and $\text{Conv2D}_{1} $ is a 1x1 convolutional layer. The excitation map $\mathbf{F}_{\text{ex}} $ is then computed as:
\begin{equation}
\mathbf{F}_{\text{ex}} = \sigma(\text{Conv2D}_{2}(\text{ReLU}(\mathbf{F}_{\text{sq}}))),
\end{equation}
where $\text{Conv2D}_{2} $ is another 1x1 convolutional layer. $\text{ReLU} $ is the rectified linear unit activation function.  $\sigma $ is the sigmoid activation function. 
Finally, the output of the channel attention mechanism $\mathbf{F}_{\text{ch}} $ is obtained by:
\begin{equation}
\mathbf{F}_{\text{ch}} = \mathbf{F} \cdot \mathbf{F}_{\text{ex}}.
\end{equation}

To efficiently merge features from branches with varying patch dimensions, we propose a cross-branch attention mechanism inspired by CrossViT \cite{chen2021crossvit}. Unlike CrossViT, which focuses on integrating CLS tokens and patch tokens for image classification, our approach facilitates direct interaction between patch tokens across scales. This concept is specifically designed for no-reference IQA (NR-IQA), where capturing both fine-grained and coarse distortions at multiple scales is critical. The mechanism is defined as follows:
\begin{equation}
\mathbf{F}_{\text{cross}} = \text{softmax}\left(\frac{\mathbf{Q}_{s}\mathbf{K}_{l}^T}{\sqrt{d_k}}\right) \mathbf{V}_{l} + \text{softmax}\left(\frac{\mathbf{Q}_{l}\mathbf{K}_{s}^T}{\sqrt{d_k}}\right) \mathbf{V}_{s},
\end{equation}
where $\mathbf{Q}_{s}$, $\mathbf{K}_{s}$, and $\mathbf{V}_{s}$ are the query, key, and value matrices derived from the short-branch features, and $\mathbf{Q}_{l}$, $\mathbf{K}_{l}$, and $\mathbf{V}_{l}$ are the corresponding matrices derived from the long-branch features. The softmax operation provides cross-attention weights, allowing one branch to focus on the other's features, resulting in successful feature integration across branches.

\begin{table*}[htbp]
\caption{Comparison of PLCC and SROCC for our method and state-of-the-art NR-IQA algorithms on benchmark datasets, with highest, second highest and third highest results marked in red, orange, and blue, respectively.}
\centering
\small  
\begin{tabular}{|c|cc|cc|cc|cc|cc|}
    \hline
    \multirow{2}{*}{\textbf{Method}} & \multicolumn{2}{c|}{\textbf{KONIQ-10k}} & \multicolumn{2}{c|}{\textbf{LIVE}} & \multicolumn{2}{c|}{\textbf{LIVE-C}} & \multicolumn{2}{c|}{\textbf{CSIQ}} & \multicolumn{2}{c|}{\textbf{Average}} \\
    \cline{2-11}
    & \textbf{PLCC} & \textbf{SROCC} & \textbf{PLCC} & \textbf{SROCC} & \textbf{PLCC} & \textbf{SROCC} & \textbf{PLCC} & \textbf{SROCC} & \textbf{PLCC} & \textbf{SROCC} \\
    \hline
    DIIVINE \cite{saad2012blind} & 0.558 & 0.546 & 0.908 & 0.892 & 0.591 & 0.588 & 0.776 & 0.804 & 0.708 & 0.708 \\
    BRISQUE \cite{mittal2012no} & 0.685 & 0.681 & 0.940 & 0.929 & 0.629 & 0.629 & 0.748 & 0.812 & 0.751 & 0.763 \\
    MEON \cite{ma2017end} & 0.628 & 0.611 & 0.955 & 0.951 & 0.710 & 0.697 & 0.942 & 0.922 & 0.809 & 0.795 \\
    DBCNN \cite{zhang2018blind} & 0.884 & 0.875 & \textcolor{orange}{0.971} & \textcolor{orange}{0.969} & 0.869 & \textcolor{blue}{0.869} & \textcolor{orange}{0.959} & \textcolor{orange}{0.946} & 0.921 & \textcolor{blue}{0.914} \\
    TRIQ \cite{you2021transformer} & \textcolor{blue}{0.903} & 0.892 & 0.965 & 0.949 & 0.861 & 0.845 & 0.838 & 0.828 & 0.892 & 0.879 \\
    HyperIQA \cite{su2020blindly} & 0.885 & 0.872 & 0.966 & 0.962 & 0.842 & 0.844 & 0.942 & 0.923 & 0.909 & 0.900 \\
    TreS \cite{golestaneh2022no} & \textcolor{orange}{0.918} & \textcolor{orange}{0.915} & \textcolor{blue}{0.968} & \textcolor{orange}{0.969} & \textcolor{blue}{0.877} & {0.846} & 0.942 & 0.922 & \textcolor{blue}{0.926} & {0.913} \\
    MAMIQA \cite{yu2023mamiqa} & \textcolor{red}{0.937} & \textcolor{red}{0.926} & \textcolor{red}{0.981} & \textcolor{red}{0.981} & \textcolor{orange}{0.895} & \textcolor{orange}{0.874} & \textcolor{red}{0.972} & \textcolor{red}{0.962} & \textcolor{red}{0.946} & \textcolor{red}{0.936} \\
    Ours & \textcolor{blue}{0.903} & \textcolor{blue}{0.909} & \textcolor{blue}{0.968} & \textcolor{blue}{0.964} & \textcolor{red}{0.903} & \textcolor{red}{0.895} & \textcolor{blue}{0.945} & \textcolor{blue}{0.925} & \textcolor{orange}{0.928} & \textcolor{orange}{0.923} \\
    \hline
\end{tabular}
\label{tab:performance}
\end{table*}

MS-SCANet lowers its computational complexity to $O(N_w^2 \cdot d)$,  leveraging window-based attention, fewer patches of larger size (16x16 and 32x32), and a smaller embedding dimension (256 compared to 768). In contrast, ViT's global attention scales quadratically at $O(N^2 \cdot d)$, where $N$ is the total number of patches. So, MS-SCANet is more efficient, requiring approximately 14.7M FLOPs per token, compared to SwinT (71.8M FLOPs/token), TRIQ (92.9M FLOPs/token), and ViT (185.7M FLOPs/token). The use of window-based attention not only limits interactions but also improves efficiency, exceeding models like SwinT and TRIQ, which, despite their task-specific optimizations, exhibit higher computational costs.

\section{Proposed Loss Functions}

To encourage cohesive feature learning across various branches, we propose two consistency loss functions: Cross-Branch Consistency Loss and the Adaptive Pooling Consistency Loss. The primary objective of the Cross-Branch Consistency Loss (CB Loss) is to maintain feature uniformity between branches by minimizing the mean squared error (MSE) between branch features, defined as follows:
\begin{equation}
\mathcal{L}_{CB} = \alpha \cdot \text{MSE}(\mathbf{F}_{s}, \mathbf{F}_{l}),
\end{equation}
where $\mathbf{F}_{s}$ and $\mathbf{F}_{l}$ denote the features from the short and  long branches, respectively, and $\alpha$ acts as a weighting coefficient.

\begin{figure}[tb]
    \centering
    \subfloat[KoNIQ-10k]{\includegraphics[width=0.23\textwidth]{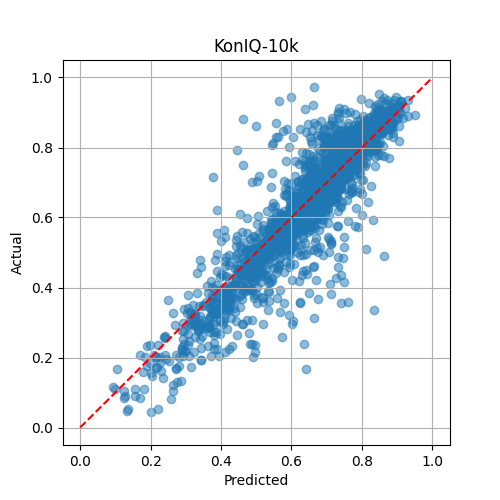}\label{fig:scatter_koniq}}
    \hfill
    \subfloat[LIVE]{\includegraphics[width=0.23\textwidth]{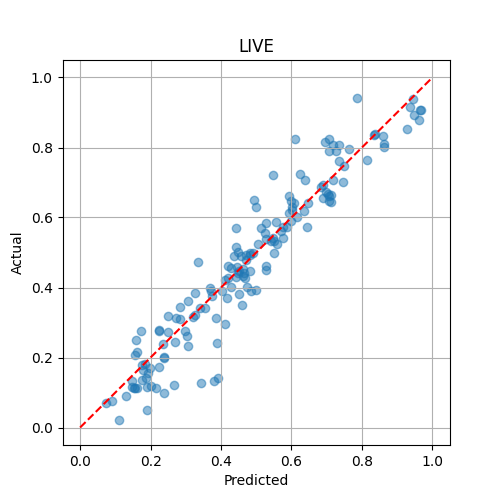}\label{fig:scatter_live}}
    \\
    \subfloat[LIVE-C]{\includegraphics[width=0.23\textwidth]{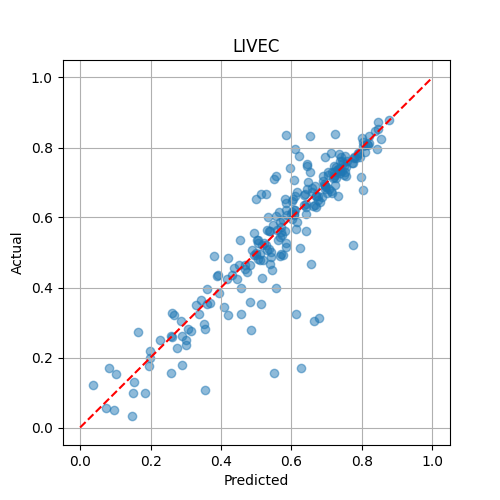}\label{fig:scatter_livec}}
    \hfill
    \subfloat[CSIQ]{\includegraphics[width=0.23\textwidth]{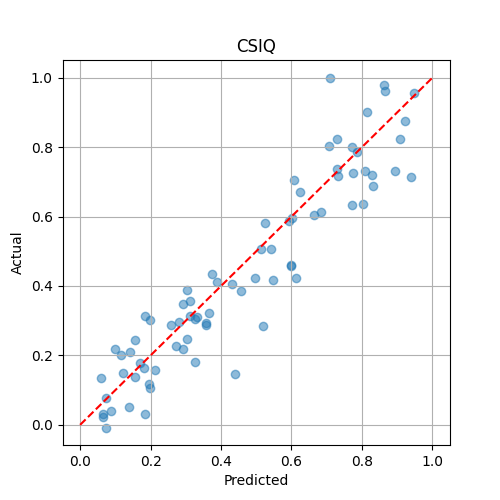}\label{fig:scatter_csiq}}
    \caption{Scatter plots and linear regression fit lines of our proposed MS-SCANet method on benchmark datasets.}
    \label{fig:scatter_plots}
\end{figure}

The Adaptive Pooling Consistency Loss (AP Loss) is designed to standardize feature maps across various scales, ensuring uniform feature integration during the downsampling phase. Unlike linear transformation and bilinear interpolation methods \cite{he2015}, which can introduce artifacts and distort the feature space during rescaling, adaptive pooling maintains the structural integrity of the features by adapting the pooling process to the content of the feature map while maintaining its quality. This loss function  is defined as:
\begin{equation}
\mathcal{L}_{AP} = \beta \cdot \text{MSE}(\mathbf{F}_{orig}, \mathbf{F}_{pool}),
\end{equation}
where $\mathbf{F}_{orig}$ and $\mathbf{F}_{pool}$ are the features before and after pooling, respectively, with $\beta$ as the associated weighting factor. Initially intended for quality evaluation, AP Loss retains spatial relationships in object detection, while CB Loss enhances scale uniformity for semantic segmentation.

The L1 loss function, also known as the mean absolute error (MAE), calculates the absolute difference between the predicted and ground truth values. It prioritizes sparsity and is less prone to outliers than other loss functions, such as L2 loss.  By minimizing L1 loss, the model aims to reduce absolute errors between anticipated and actual image quality scores, resulting in more stable and predictable performance. The total training loss is a combination of the core L1 loss and the consistency loss functions. 
\begin{equation}
\mathcal{L}_{total} = \mathcal{L}_{L1} +  \mathcal{L}_{CB} + \mathcal{L}_{AP}.
\end{equation}
We implemented the model using PyTorch, training on an NVIDIA RTX 4000 Ada GPU with an initial learning rate of $1 \times 10^{-4}$. A cosine annealing schedule and a batch size of 16 over 600 epochs was used for learning rate adjustments, with consistency loss weights $\alpha$ and $\beta$ set to 0.5. 

\begin{table*}[htbp]  
\centering
\caption{Ablation Study Results: Attention Mechanisms. The best results are in bold.}
\label{tab:ablation_study_attention}
\small  
\begin{tabular}{|l|cc|cc|cc|cc|}
    \hline
    \multirow{2}{*}{\textbf{Configuration}} & \multicolumn{2}{c|}{\textbf{LIVE}} & \multicolumn{2}{c|}{\textbf{KonIQ-10k}} & \multicolumn{2}{c|}{\textbf{LIVE-C}} & \multicolumn{2}{c|}{\textbf{CSIQ}} \\
    \cline{2-9}
    & \textbf{PLCC} & \textbf{SROCC} & \textbf{PLCC} & \textbf{SROCC} & \textbf{PLCC} & \textbf{SROCC} & \textbf{PLCC} & \textbf{SROCC} \\
    \hline
    Single Branch both attention & 0.939 & 0.935 & 0.887 & 0.875 & 0.865 & 0.854 & 0.922 & 0.910 \\
    Single branch Spatial Attention & 0.955 & 0.950 & 0.846 & 0.856 & 0.873 & 0.862 & 0.929 & 0.918 \\
    Single branch Channel Attention & 0.951 & 0.948 & 0.844 & 0.844 & 0.869 & 0.860 & 0.927 & 0.916 \\
    Multi-branch dual attention (Proposed) & \textbf{0.968} & \textbf{0.964} & \textbf{0.903} & \textbf{0.909} & \textbf{0.903} & \textbf{0.895} & \textbf{0.937} & \textbf{0.925} \\
    \hline
\end{tabular}
\end{table*}

\begin{table*}[htbp]  
\centering
\caption{Ablation Study Results: Different Loss Functions. The best results are in bold.}
\label{tab:ablation_study}
\small  
\begin{tabular}{|l|cc|cc|cc|cc|}
    \hline
    \multirow{2}{*}{\textbf{Configuration}} & \multicolumn{2}{c|}{\textbf{LIVE}} & \multicolumn{2}{c|}{\textbf{KonIQ-10k}} & \multicolumn{2}{c|}{\textbf{LIVE-C}} & \multicolumn{2}{c|}{\textbf{CSIQ}} \\
    \cline{2-9}
    & \textbf{PLCC} & \textbf{SROCC} & \textbf{PLCC} & \textbf{SROCC} & \textbf{PLCC} & \textbf{SROCC} & \textbf{PLCC} & \textbf{SROCC} \\
    \hline
    L1 Loss Only (Baseline)         & 0.968 & 0.964 & 0.903 & 0.909 & 0.903 & 0.895 & 0.937 & 0.925 \\
    L1 + CB Loss                    & 0.971 & 0.969 & 0.910 & 0.916 & 0.907 & 0.902 & 0.942 & 0.930 \\
    L1 + AP Loss                    & 0.969 & 0.969 & 0.915 & 0.917 & 0.908 & 0.899 & 0.940 & 0.928 \\
    L1 + CB Loss + AP Loss (Full Model) & \textbf{0.972} & \textbf{0.973} & \textbf{0.921} & \textbf{0.923} & \textbf{0.909} & \textbf{0.907} & \textbf{0.945} & \textbf{0.933} \\
    \hline
\end{tabular}
\end{table*}

\begin{figure}[htbp]
    \centering
    \includegraphics[width=0.45\textwidth]{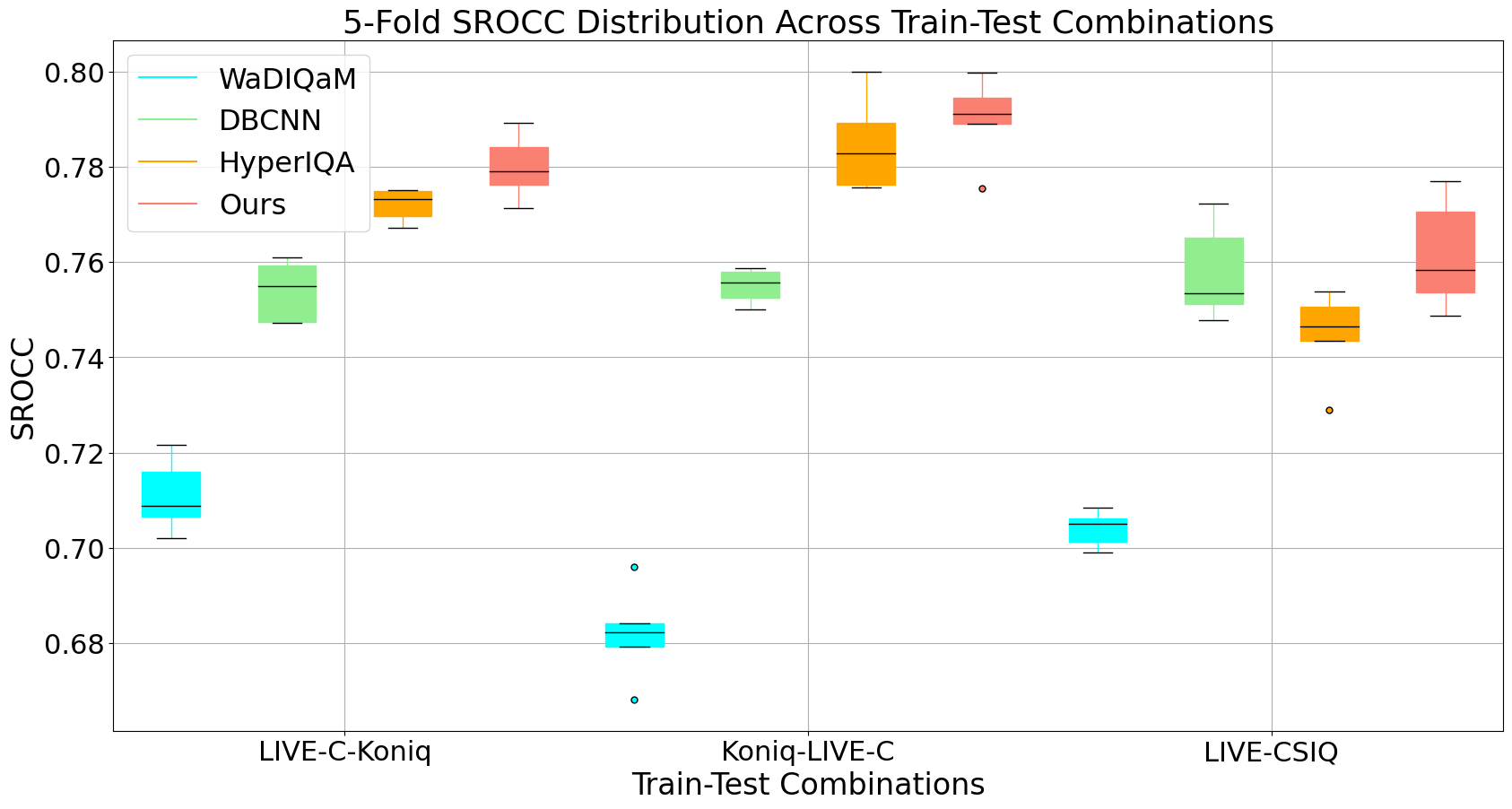}  
    \caption{Box plot for SROCC for cross-dataset combination for different models: WaDIQaM \cite{bosse2018deep}, DBCNN \cite{zhang2018blind}, HyperIQA \cite{su2020blindly}, and MS-SCANet across three train-test combinations. The label on the \textit{x}-axis, dataset1-dataset2, indicates that training was conducted on dataset1 while testing was performed on dataset2.  }
    \label{cv_plot}
\end{figure}

\section{Experimental Results}
\label{sec:results}
We evaluated the performance of our MS-SCANet model on four popular benchmark datasets: KonIQ-10k \cite{lin2019koniq-10k}, LIVE \cite{sheikh2006statistical}, LIVE Challenge \cite{ghadiyaram2016live}, and CSIQ \cite{chandler2010most}. KonIQ-10k includes 10,073 images with distortions such as noise, blur, compression artifacts, and blocking. The LIVE dataset contains 779 images and visual artifacts such as JPEG2000, JPEG, Gaussian noise, and blur. The LIVE Challenge dataset contains 1,162 images with real-world distortions collected from user-generated content. Finally, the CSIQ dataset contains 866 images with various synthetic distortions, including additive noise, JPEG compression, etc. The assessment followed standard IQA protocols, using Pearson's Linear Correlation Coefficient (PLCC) and Spearman's Rank-Order Correlation Coefficient (SROCC) to measure model accuracy. 

We compared MS-SCANet with several state-of-the-art NR-IQA models. The compared approaches are classified as follows: DIVINE and BRISQUE under natural scene statistics methods; MEON, DBCNN, HyperIQA, and MAMIQA under CNN-based methods; and TRIQ and TreS under transformer-based methods. As depicted in Table \ref{tab:performance}, MS-SCANet consistently ranked within the top three metrics in terms of PLCC and SROCC (best results are presented in \textit{Red, Orange and Blue}, respectively). Empirical analysis shows that a 6$\times$6 window size with a 256 embedding dimension provides a balanced trade-off between PLCC and SROCC. The findings demonstrated robust performance on the LIVE Challenge dataset, an especially challenging dataset, where it attained the highest correlation. Figure \ref{fig:scatter_plots} shows the MS-SCANet scatter plots in all datasets. These results indicate the capability of MS-SCANet to handle synthetic and natural visual distortions such as noise, blur, compression artifacts, and blocking. The model's dual-branch design and cross-branch attention mechanism enable it to process fine and coarse image characteristics, thus detecting local and global artifacts. To evaluate the generalization capability of MS-SCANet across multiple datasets, we used a 5-fold cross-dataset validation procedure where we trained in one dataset and tested in another. Figure \ref{cv_plot} shows a box plot with the distribution of SROCC values in various train-test scenarios, where the label in the \textit{x}-axis indicates the train-test combination. MS-SCANet demonstrated strong performance and low variability, especially when trained on KonIQ-10k and tested on LIVE-C, indicating robustness to various distortions.

We performed an ablation analysis to determine the contribution of each MS-SCANet component to the overall performance. Table \ref{tab:ablation_study_attention} compares models utilizing spatial attention, channel attention, and dual attention with the proposed dual-attention MS-SCANet. The multiscale dual attention configuration consistently outperformed other configurations, underscoring the importance of integrating attention mechanisms across scales to capture local and global features. In addition, we analyzed the impact of consistency loss functions individually and in combination. Table \ref{tab:ablation_study} shows that the combination of cross-branch consistency loss (CB-Loss) and adaptive pooling consistency loss (AP-Loss) significantly enhanced performance by improving feature integration during downsampling.

\section{Conclusion}
We introduce MS-SCANet, a BIQA Multi-Scale Dual Attention Transformer model with a dual-branch design to capture fine-grained and coarse-grained features, enhanced by spatial and channel attention mechanisms. The cross-branch attention mechanism integrates feature representations from different scales effectively. Additionally, two new consistency loss strategies, cross-branch consistency and adaptive pool consistency, improve feature learning and multiscale fusion. Extensive experiments on four benchmark datasets show that MS-SCANet outperforms current methods, with cross-dataset evaluations confirming its strong generalizability. In future work, we plan to explore further enhancements in computational efficiency and the potential applicability of the proposed model to video quality assessment tasks.
\bibliographystyle{IEEEtran}
\bibliography{refs}

\end{document}